\documentclass[%
reprint,
amsmath,amssymb,
aps,
]{revtex4-2}

\usepackage[english]{babel}
\usepackage{graphicx}
\usepackage{dcolumn}
\usepackage{bm}

\begin{document}

\preprint{APS/123-QED}

\title{
Auditory frequency analysis as an active dissipative process
}
\thanks{
Auditory frequency analysis as an active dissipative process
}

\author{Yasuki Murakami}
\altaffiliation{Faculty of Design, Kyushu University, 4-9-1, Minami-ku, Fukuoka 815–8540, Japan.}
\email{murakami@design.kyushu-u.ac.jp}
\date{\today}

\begin{abstract}
An active dissipative process organizes auditory frequency analysis in the mammalian cochlea. A minimal active beam model reveals that a spatially varying viscous coupling operator, $\partial_{xx}\kappa\partial_{xx}$, generates dissipative forces with wave--like propagation. Local energy injection and spatial redistribution compete to govern the dynamics. This balance enables the quantitative reproduction of four key features: sharp tuning, high gain, compression, and spontaneous otoacoustic emissions. Hearing thereby belongs to a broad class of nonequilibrium pattern-forming systems.
\end{abstract}
%
\maketitle

Many physical systems operate far from thermal equilibrium, continuously exchanging energy with their environment. These driven–dissipative media exhibit macroscopic organization and function resulting from the interplay between energy injection, dissipation, and spatial coupling. Examples include lasers~\cite{Kogelnik1972,Feng2014}, pattern-forming fluids, chemical oscillations, and active matter~\cite{Cross1993}. In such systems, dissipation is not merely an inevitable loss but serves as an essential factor in organizing stability, mode selection, and information flow.

The cochlea in mammals is a nonequilibrium system that performs real-time frequency analysis with high sensitivity and dynamic range. Conventional cochlear mechanics fall into two main categories: transmission-line models describing wave propagation along the basilar membrane~\cite{Peterson1950, Neely1986, Zweig1991}, and local oscillator models using Hopf bifurcation dynamics~\cite{Choe1998, Camalet2000, Eguiluz2000,Duke2003}. Shera~\cite{Shera2003, Shera2007} bridged these perspectives, showing an analogy between cochlear amplification and laser physics and suggesting that the cochlea is an active medium far from equilibrium. Later studies found that viscosity contributes to cochlear mechanics~\cite{Prodanovic2019,Deloche2025,Shokrian2025,Lee2025}. The exact role of viscous dissipation in organizing frequency selectivity remains unclear. Optical coherence tomography (OCT) reveals that velocities vary spatially in the intracochlear space~\cite{Cooper2018, Dewey2019, Olson2025}. The OCT experiments point to nanoscale viscous interactions.

A viewpoint rooted in modern driven–dissipative physics is adopted in which dissipation plays an essential organizational role in auditory frequency analysis. A minimal viscously coupled active beam model shows that frequency selectivity arises from competition between local energy injection and spatial energy redistribution. Dissipation thereby organizes spectral responses through the structure of the operator, placing the cochlea within the broader class of driven–dissipative systems.

\begin{figure*}[t]
\centering
\begin{minipage}{0.64\linewidth}
\includegraphics[width=\linewidth]{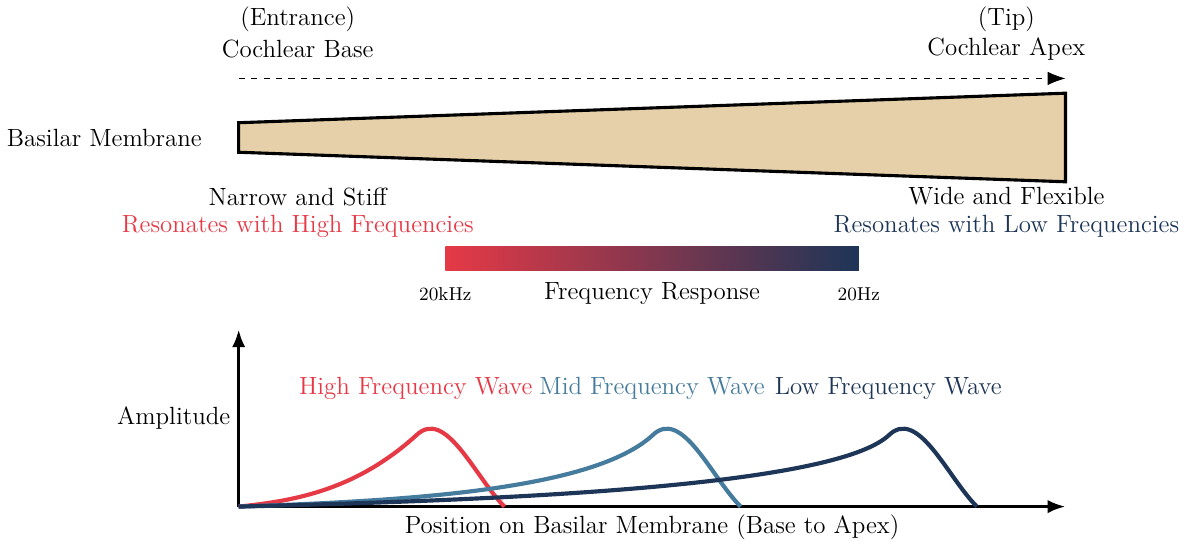}
\end{minipage}
\begin{minipage}{0.35\linewidth}
\centering
\includegraphics[width=\linewidth]{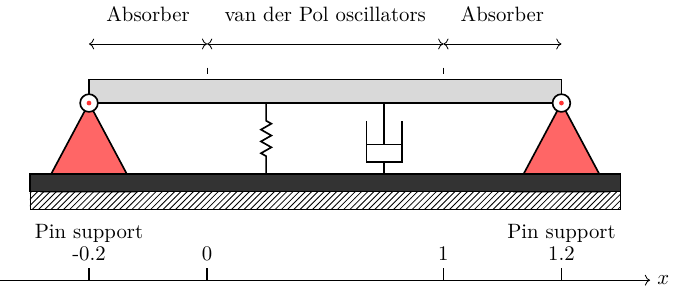}
\includegraphics[width=0.8\linewidth]{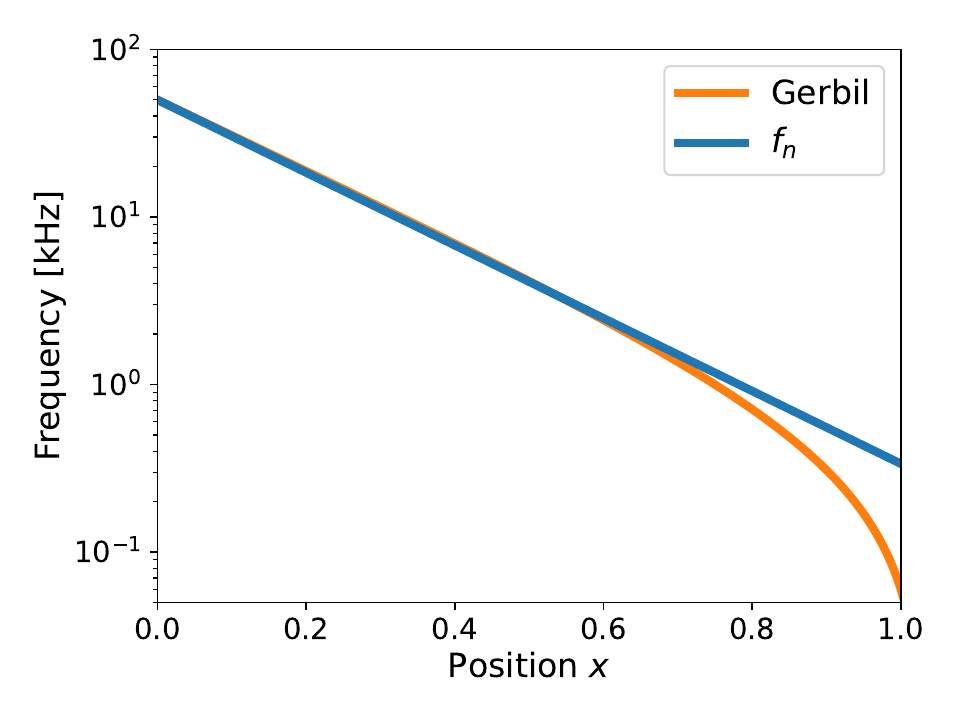}
\end{minipage}
\caption{Physiologically grounded schematic of the model.
(Left) Key anatomical and functional features of the mammalian cochlea motivating the model.
(Right, top) Minimal active beam model with spatially varying viscous coupling, including a central active region ($\mu>0$) and passive absorbing boundaries.
(Right, bottom) Physiological tonotopy implemented using the Greenwood frequency–position function $f(x)$ \cite{Greenwood1990}.
}

\label{fig1}
\end{figure*}

The cochlea is a fluid–structure interaction system. Basilar membrane (BM) motion and cochlear fluids are bidirectionally coupled. We account for this coupling through viscous terms and model the BM as a viscoelastic beam with active feedback. The model is based on three assumptions. (i) The longitudinal Young’s modulus is negligible \cite{Gummer1981, Liu2008}. (ii) The BM shows transverse viscoelasticity, approximated by spring–damper elements \cite{Gummer1981, Liu2008}. (iii) Active amplification is represented by negative damping using Hopf bifurcation dynamics and a van der Pol–type formulation \cite{Choe1998, Camalet2000, Eguiluz2000,Brownell1985,Ashmore1987,Jaramillo1993,Martin2000,Liberman2002,Kennedy2005}.

Fig.~\ref{fig1} shows the model based on these assumptions. The BM dynamics based on the Euler-Bernoulli hypothesis follow:

 \begin{equation}
     \partial_{tt} w - \mu (1-w^2)\partial_t w + \partial_{xx}\!\left(\kappa \partial_{xxt} w\right) + \omega_n^2 w = F ,
\label{eq:model}
 \end{equation}
In this model, $\mu=\mu_0 s(x)c(x)$, $\kappa=\kappa_0 s(x)$, $\omega_n=\omega_0 s(x)$, and $s(x)=e^{-\alpha x}$. The function $c(x)$ smooths the transition from the active region to the absorbing boundaries. Here $w$ is the BM displacement, $\mu$ the active gain parameter, $\kappa$ the longitudinal viscous coupling strength, $\omega_n$ the local natural frequency (Fig.~\ref{fig1}), and $F$ the external forcing. All variables and parameters are expressed in dimensionless form, providing a consistent basis for analyzing cochlear dynamics.

For small amplitudes ($w\ll1$), Eq.~(\ref{eq:model}) reduces to
\begin{equation}
    \partial_{tt} w + \mathcal{L}\partial_t w + \omega_n^2 w = F , \label{eq:model_LIN}
\end{equation}
with
\begin{equation}
    \mathcal{L}=-\mu+\partial_{xx}\kappa\partial_{xx}. \label{eq:L}
\end{equation}
A key feature is that the spatially varying viscosity $\kappa(x)$ is inside the differential operator. This allows viscous forces to propagate instead of acting only as local damping. Along with the local negative damping $-\mu$, this structure defines an active dissipative medium.

To elucidate the spectral properties of the dissipative coupling, we examine its representation in wavenumber $k$ space. Fourier transformation yields
\begin{equation}
\mathcal{F}!\left[\partial_{xx}\kappa\partial_{xx}\right] \propto k^2-\alpha^2+2i\alpha k .
\label{eq:fourier}
\end{equation}
Equation \ref{eq:fourier} follows directly from an exact Fourier transform of Eq. (\ref{eq:L}) with spatially varying $\kappa(x) = \kappa_0 e^{-\alpha x}$, without further approximation.
The real part $ k^2- \alpha^2$ causes wavenumber-dependent attenuation. The imaginary part $2\alpha k$ produces a graded phase shift, leading to wave-like propagation. When $\alpha>0$, describing graded material properties, this term breaks spatial symmetry and enables directional propagation of dissipative modes. For a homogeneous case ($\alpha=0$), the imaginary part disappears, leaving only diffusive coupling.

Absorbing boundaries are used to suppress reflections; the main results are insensitive to the specific boundary implementation. This is implemented by smoothly reducing the activity parameter $\mu(x)$ from positive values in the active region to negative values near the boundaries using the weight function $c(x)$, consistent with the absence of active elements at the basal and apical extremes of the basilar membrane.

\begin{figure*}[ht]
\centering
\begin{minipage}{0.32\linewidth}
\centering
\includegraphics[width=\linewidth]{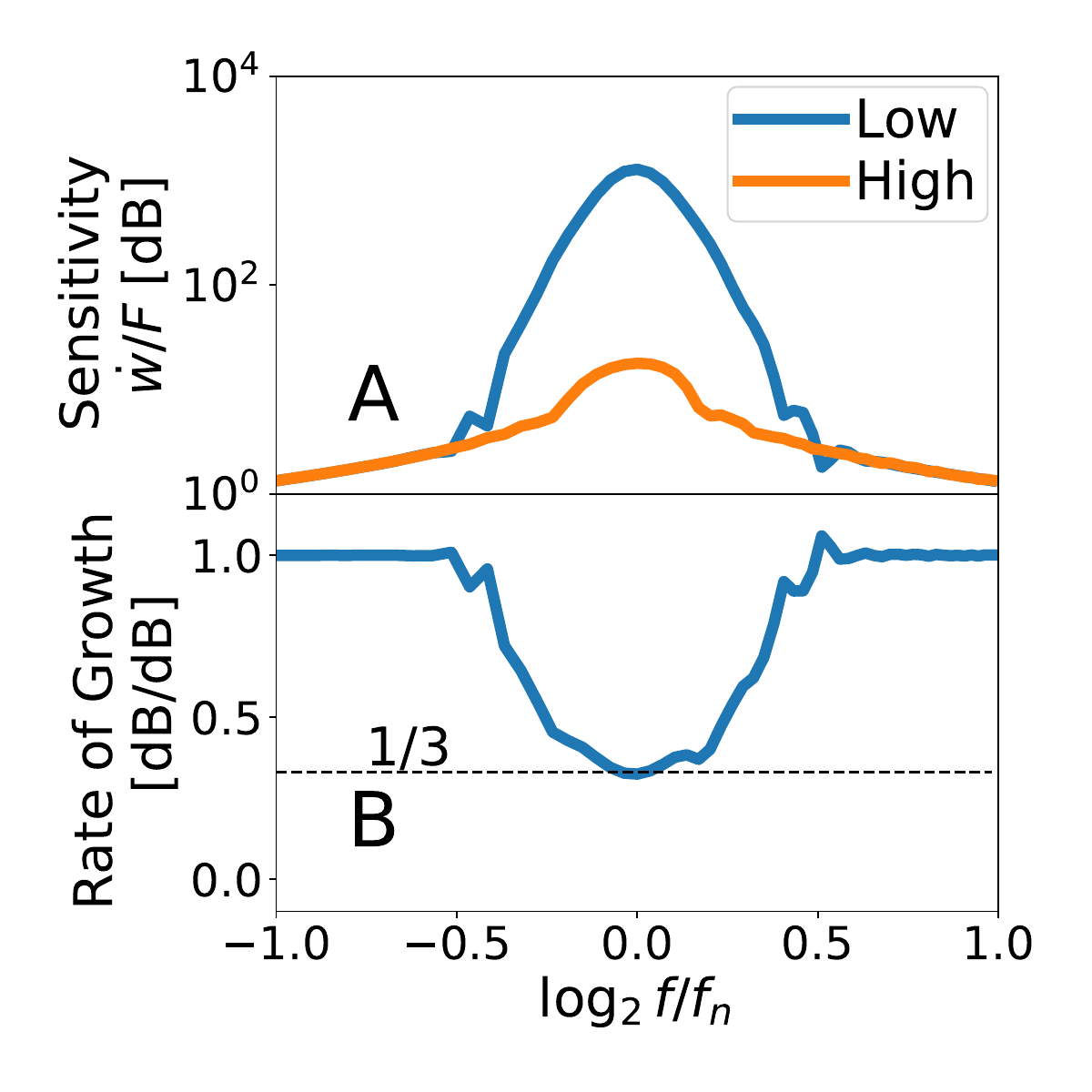}
\end{minipage}
\begin{minipage}{0.32\linewidth}
\includegraphics[width=\linewidth]{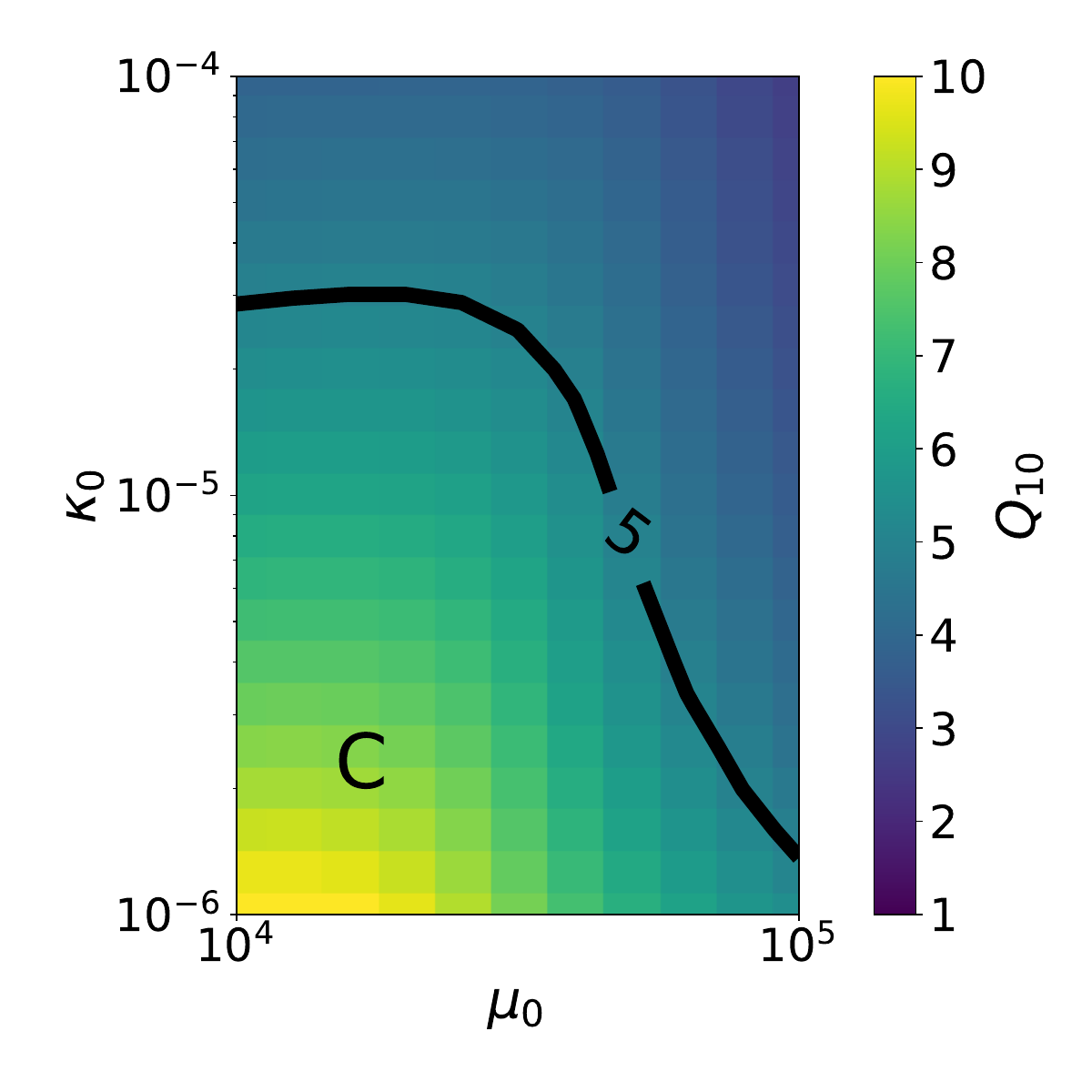}
\end{minipage}
\begin{minipage}{0.32\linewidth}
\includegraphics[width=\linewidth]{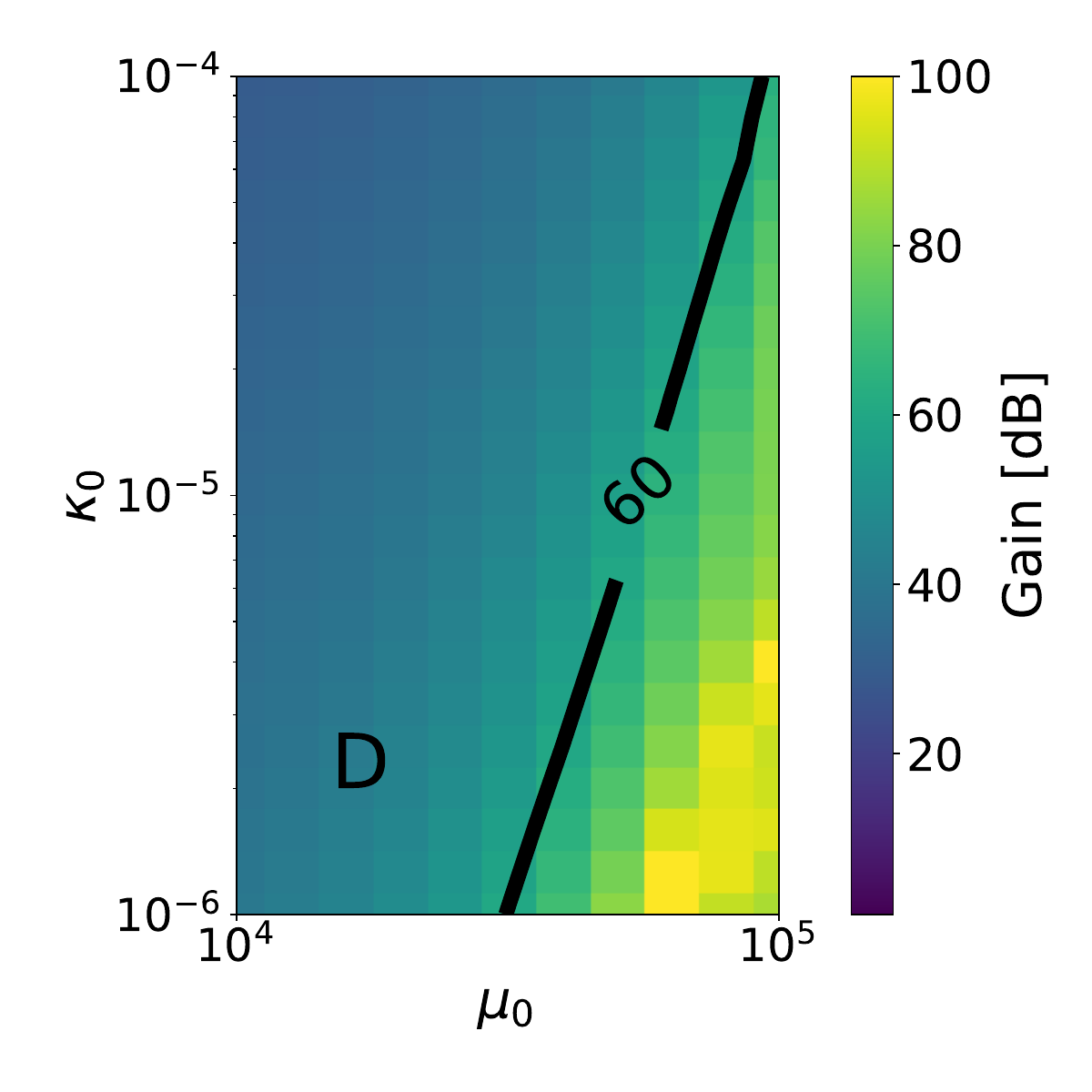}
\end{minipage}
\caption{
(A) Sensitivity for different input levels with $\mu_0=5\times10^4$ and $\kappa_0=10^{-5}$.
(B) Rate of growth (ROG).
(C) Quality factor $Q_{10}$ in the $(\mu_0,\kappa_0)$ parameter space.
(D) Mechanical amplification gain in the same parameter space.
Black contours in  (C) and (D) indicate physiologically relevant ranges.
}
\label{fig2}

\end{figure*}

The properties of the dissipative operator and their implications for frequency selectivity are first examined. Numerical responses of the full model are then presented to illustrate these effects. The model exhibits level-dependent sensitivity characteristic of cochlear responses (Fig.~\ref{fig2}A): sensitivity is high at low input levels and decreases at high levels, while responses far from the characteristic frequency remain weak and largely level-independent. Frequency selectivity ($Q_{10}\approx5$) and mechanical gain ($\sim60$ dB) fall within experimentally observed ranges \cite{Sellick1982,Rhode2007,Ren2011,Cooper2018}. The input–output relation shows compressive nonlinearity with slopes of order $\sim1/3$ dB/dB (Fig.~\ref{fig2}B), consistent with measurements \cite{Rhode2007,Alonso2025}. Varying the activity parameter $\mu_0$ and viscous coupling strength $\kappa_0$ reveals their distinct roles (Fig.~\ref{fig2}C,D): smaller $\kappa_0$ produces sharper tuning (larger $Q_{10}$), whereas larger $\mu_0$ increases amplification gain. Thus, frequency selectivity and gain are jointly controlled by the balance between local activity and nonlocal viscous coupling.

\begin{figure*}[t]
\centering
\begin{minipage}{0.49\linewidth}
\includegraphics[width=\linewidth]{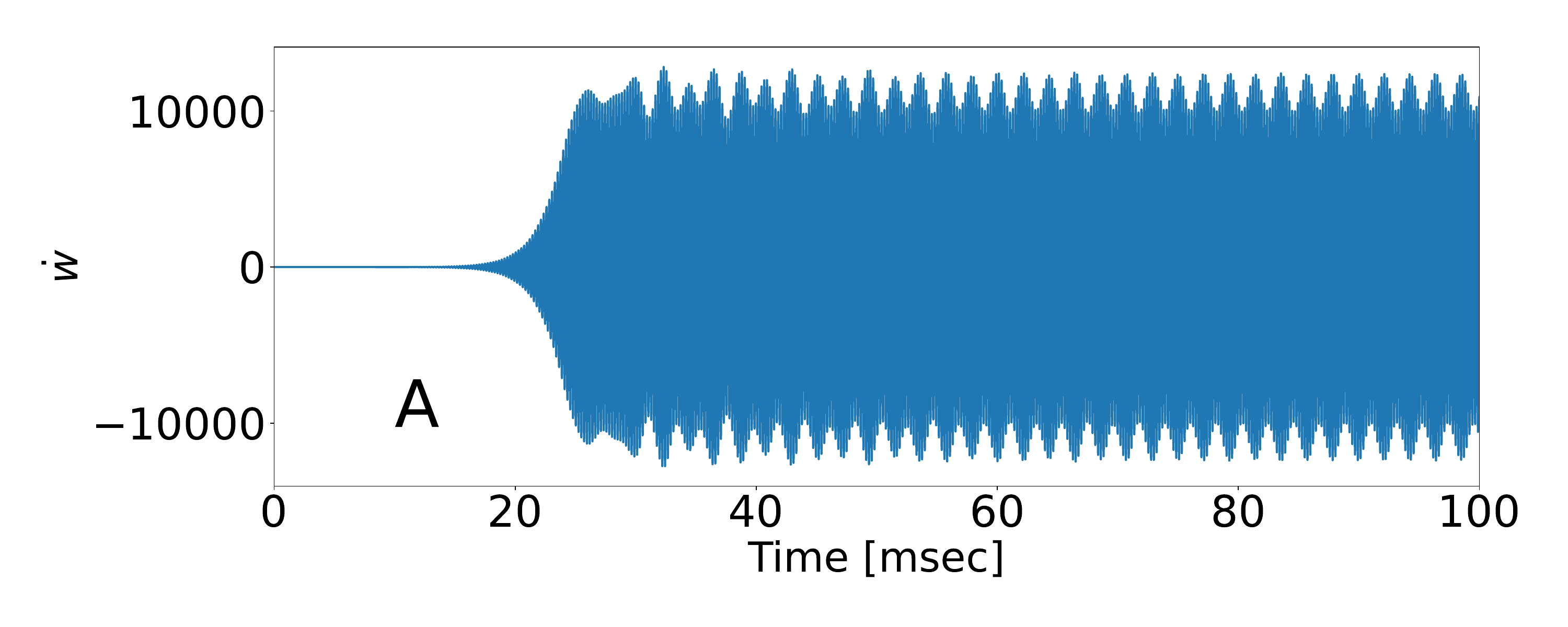}
\end{minipage}
\begin{minipage}{0.49\linewidth}
\includegraphics[width=\linewidth]{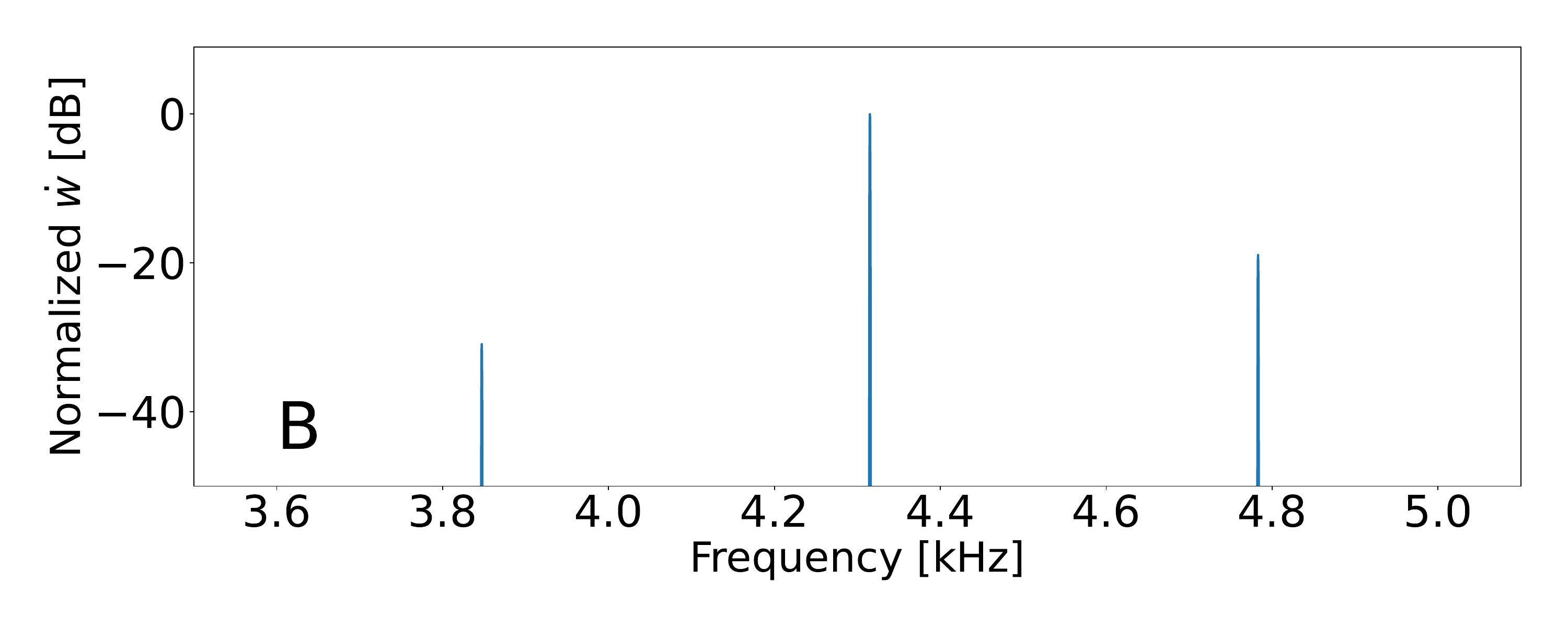}
\end{minipage}
\caption{
SOAEs generated by a localized reduction (notch) in the activity and viscous-coupling profiles.
(A) Temporal response showing sustained spontaneous oscillations with amplitude modulation.
(B) Power spectral density exhibiting a dominant emission peak accompanied by regularly spaced sidebands.
}
\label{fig3}
\end{figure*}

Spontaneous otoacoustic emissions (SOAEs) are generated by introducing a localized reduction (notch) in the activity and viscous-coupling profiles. At the notch, $\mu$ and $\kappa$ are set to zero at a discrete spatial location. This perturbation produces localized instabilities that model physiological impedance variations implicated in SOAE generation \cite{Shera2003,Shera2007}. SOAEs thus arise naturally from the same dissipative-wave mechanism underlying amplification and frequency selectivity. The temporal response exhibits sustained spontaneous oscillations with amplitude modulation (Fig.~\ref{fig3}A). The corresponding power spectral density (Fig.~\ref{fig3}B) shows a dominant emission peak accompanied by regularly spaced sidebands, consistent with experimental observations \cite{Talmadge1993,Braun1997,Shera2022}. The model thus reproduces the characteristic spectral organization of SOAEs.

The cochlea has long been recognized as an active sensory organ, and the laser-amplifier analogy~\cite{Shera2003,Shera2007} suggests that it operates as an active medium far from equilibrium in which dissipation is unavoidable and functionally relevant. Dissipation plays an essential organizational role in auditory frequency analysis, as revealed by the operator $\mathcal{L}$ in Eq. (\ref{eq:L}) which encodes a fundamental competition between local energy injection through negative damping and spatial energy redistribution through viscous coupling. It is this balance that enables the simultaneous emergence of sharp frequency selectivity and large mechanical gain. The quantitative reproduction of compressive nonlinearity, high amplification, sharp tuning, and SOAEs demonstrates that dissipation is not merely compatible with cochlear function but constitutes a constructive element underlying it.

Local Hopf bifurcation models have successfully described cochlear amplification and compression~\cite{Choe1998, Camalet2000, Eguiluz2000,Duke2003}. Here the emphasis is shifted to the spatial organization of dissipation, which governs frequency selectivity through the structure of a dissipative operator. The van der Pol nonlinearity provides local Hopf dynamics essential for amplification; the dissipative operator $\mathcal{L}$ organizes these local elements into a spatially coherent frequency analyzer.

The present viscous beam model should be regarded as an effective description in which fluid–structure interaction effects are incorporated into the viscous coupling term.~\cite{Prodanovic2019,Deloche2025,Shokrian2025,Lee2025}. At the low Reynolds numbers relevant to the cochlea, viscous forces dominate over inertia, allowing fluid influences to be represented without explicitly solving pressure fields~\cite{Purcell1977}. The ability of this minimal model to reproduce compression, gain, sharp tuning, and SOAEs indicates that essential cochlear physics is preserved despite the abstraction. The models incorporating viscous fluid–structure interactions \cite{Prodanovic2019,Deloche2025,Shokrian2025,Lee2025} provide candidate mechanisms for $\kappa(x)$; the present analysis offers theoretical support for these approaches by demonstrating that spatially graded viscous coupling is structurally essential to frequency selectivity."

The competition between local energy injection and spatial redistribution represents a universal structure shared across driven–dissipative systems~\cite{Kogelnik1972,Feng2014,Cross1993}. In laser amplifiers, for example, local gain from the active medium competes with diffraction and transverse mode coupling to determine spectral and spatial organization. In the cochlea, this competition is embodied in the biharmonic viscous operator $\partial_{xx}\kappa\partial_{xx}$, which produces wavenumber-dependent attenuation and, in the presence of an exponential gradient, directional wave propagation. This framework provides a mechanistic realization of the laser-amplifier analogy~\cite{Shera2003,Shera2007}, demonstrating that dissipation plays a dual role as both a sink and an organizing element of spectral responses.

Active dissipation thus emerges as a fundamental physical principle underlying auditory frequency analysis. Hearing is thereby identified as an active dissipative process, placing cochlear frequency analysis within the broader class of nonequilibrium pattern-forming systems.

\bibliography{PRL_ddws}

\end{document}